\documentclass[12pt,a4paper]{article}
\usepackage{ifpdf}
\ifpdf
\pdfpageattr {/Group << /S /Transparency /I true /CS /DeviceRGB>>}
\fi
\usepackage{jheppub}
\usepackage{amsmath}
\usepackage{caption}
\usepackage{subfigure}

\newcommand{\beq}{\begin{equation}}
\newcommand{\eeq}{\end{equation}}
\newcommand{\bea}{\begin{eqnarray}}
\newcommand{\eea}{\end{eqnarray}}

\begin{document}

\title{A Stochasticity Threshold in Holography and the Instability of AdS }
\author[a]{Pallab Basu,} \ 
\author[b]{Chethan Krishnan, } \
\author[a, b, c]{Ayush Saurabh } \
\emailAdd{pallabbasu@gmail.com}
\affiliation[a]{International Center for Theoretical Sciences, \\
  Indian Institute of Science Campus, \\ 
  Bangalore - 560012, \ \ India}
\emailAdd{chethan.krishnan@gmail.com}
\affiliation[b]{Center for High Energy Physics, \\
  Indian Institute of Science, \\ 
  Bangalore - 560012, \ \ India}
\affiliation[c]{School of M. A. C. E., The University of Manchester,\\ 
  Manchester - M13 9PL, \ \ United Kingdom}
\emailAdd{ayushsaurabh@hotmail.com}
\abstract{We give strong numerical evidence that a self-interacting probe scalar field in AdS, with only a few modes turned on initially, will undergo fast thermalization only if it is above a certain energetic threshold. Below the threshold the energy stays close to constant in a few modes for a very long time instead of cascading quickly. This indicates the existance of a Strong Stochasticity Threshold (SST) in holography. The idea of SST is familiar from certain statistical mechanical systems, and we suggest that it exists also in AdS gravity. This would naturally reconcile the generic non-linear instability of AdS observed by Bizon and Rostworowski, with the Fermi-Pasta-Ulam-Tsingou-like quasi-periodocity noticed recently for some classes of initial conditions. We show that our simple set-up captures many of the relevant features of the full gravity-scalar system.

}

\setcounter{tocdepth}{2}
\maketitle

\section{Introduction and Conclusion}

In this paper, we will investigate the thermalization properties of a self-interacting probe scalar field in AdS. Our goal is to get some intuition about holographic thermalization in AdS gravity. We will find that for the questions we are interested in, the scalar probe has remarkable similarities to the full gravity system.

The set up we will consider in detail is as follows. We will launch the system with only a few modes excited, and then trace its evolution as the energy migrates to other modes. We find that the qualitative behavior of the system is drastically different above and below a certain threshold energy, $E_c$. 

We study this problem, first by restricting our attention to a finite-mode truncation of the system and keeping track of the (suitably time-averaged) energy per mode as the system evolves. We find that the energy per mode remains roughly constant when the system is kicked off (with a small number of modes excited) with $E < E^T_c$. But when $E > E^T_c$ we find that there is fast thermalization, and that the energy in each mode in the truncated system approaches the same value. The transition from quasi-periodicity to fast thermalization is a very sharp one in the truncated case, even though the threshold energy and the thermalization times are dependent on the initial modes we start with.

We also investigate the self-interacting scalar probe, without truncating it to a finite number of modes. Note that now, since all infinite number of modes are present, the average energy per mode in the final thermal state should be zero if one starts with a finite energy. We find that a threshold energy, $\sim E^U_c$, exists also in this un-truncated system, above which the thermalization time falls quickly by many orders of magnitude. The behavior of the full un-truncated system is less sharp that the truncated case, reminiscent more of a crossover than a phase transition, but the phenomenon is again qualitatively very robust. In both the truncated and untruncated cases, we find that the thermalization timescales do not exhibit sharp changes as we further increase the energy above their respective $E_c$.

The above paragraphs summarize our main results in this paper: when the initial modes are of small enough amplitude, quasi-periodic behavior persists and ergodicity does not set in for a long time
, for the probe scalar in AdS. Now we turn to a brief summary of the relevance of these results as well as a discussion of some auxiliary observations. In the concluding section, we will suggest that a similar statement also holds for dynamical gravity in AdS.

Our primary motivation for studying this system is as an instructive toy model for holographic thermalization and for questions about the stability of AdS. AdS spacetimes with a reflecting boundary condition have been conjectured to be unstable at the non-linear level \cite{Bizon, BizonReview}. It was argued that generic perturbations 
evolve into black holes after multiple reflections at the boundary \cite{Bizon}. This should {\em not} be thought of as an instability of the AdS ground state, but rather as the statement that generic excited states (even low-lying ones) eventually thermalize. This question is obviously of great physical interest not just from the viewpoint of understanding dynamical black holes and horizon formation (and the various related paradoxes), but also from the perspective of understanding thermalization in the dual strongly coupled field theory. 

Even though this non-linear turbulent instability is believed to be generic, 
it has been noted recently by \cite{Buchel} 
that when only certain low-lying modes are excited, and the Einstein-(massless-)scalar system in AdS \footnote{This was the system where \cite{Bizon} had noted evidence for the non-linear instability of AdS. But, see also \cite{Dias}. A very recent analytic approach to secular term resummation in the AdS instability problem can be found in \cite{Ben}.} is allowed to evolve, the system in fact does {\em not} thermalize for long periods. While the existence of the non-linear instabiltiy in the system for low amplitude coherent initial data is robust, the claims regarding the existance of the instability when only a few low-lying modes are excited are challenged by the results of \cite{Buchel}. 
They found instead that the evolution is quasi-periodic, and suggested that this behavior is analogous to the evolution of the famous Fermi-Pasta-Ulam-Tsingou (FPUT) lattice. One version of the FPUT paradox is the observation that when a self-interacting fluctuating chain with the Hamiltonian \footnote{This Hamiltonian is referred to as the FPUT $\beta$-system.}
\begin{eqnarray}
H(q,p)=\sum_{i=1}^N \frac{1}{2} p_i^2+\frac{1}{2} (q_{i+1}-q_i)^2+\frac{\mu}{4}(q_{i+1}-q_i
)^4
\end{eqnarray}
is allowed to evolve, the system does not thermalize for a very long time when only low lying modes are excited initially. Instead it exhibits quasi-periodic behavior. $N$ here is the number of lattice sites. 

However, it is known in statistical mechanics that similar slow thermalization behavior is in fact not tied only to the FPUT lattice, but that other non-linear systems also exhibit similar behavior \cite{Pettini, Yoshimura} for low amplitudes. One such example is the quartic self-intreacting scalar on a lattice:
\bea
H(q,p)=\sum_{i=1}^N \frac{1}{2} p_i^2+\frac{1}{2} (q_{i+1}-q_i)^2+\frac{\mu}{4}q_i^4
\eea
This is the massless version of a case considered in \cite{Pettini}.
A bounded lattice with Dirichlet boundary conditions is analogous to the AdS box with reflecting boundary conditions, so one might wonder whether a self-interacting scalar field in the AdS geometry captures any of these features. As a by-product of our investigation, we will find indeed that it does: we find that the solutions are quasiperiodic for a long time, and we will also be able to reproduce the qualitative features of the relevant curves presented in \cite{Buchel} with our simple probe scalar system. 


The conclusion of our work is that we find that when the initial modes are of small enough amplitude, quasi-periodic behavior persists and ergodicity does not set in\footnote{At least during the time period of the simulation! It is expected that given enough time these systems will \cite{FPUbook} ultimately thermalize.}. 
These observations are in alignment with the idea that for many non-linear systems, there is a Strong Stochastisicty Threshold (SST) and thermalization happens only if there is enough energy (amplitude) in the modes. This remark has been made in various forms before in the statistical physics literature and some relevant papers are \cite{Chirikov, Pettini, Yoshimura}. 
If the strong stochaticity threshold result holds not just for the probe scalar, but also for the backreacting gravity system, the observations made in \cite{Bizon} and \cite{Buchel} as well as the results of this paper can be seen to be consistent with this same general phenomenon, namely the existence of the SST in holographic systems; see the discussion in our section \ref{sstsec}. Indeed, this can be seen as the holographic manifestation of the Strong Stochastic Threshold (HSST): for small enough energies in the low-lying modes, one has quasi-periodic behavior and for high enough energy the behavior changes qualitatively. 

In the rest of the paper, we first introduce the systems under consideration and write down the equations of motion (section 2), describe our numerical set up and present the various numerical results that justify the claims in the abstract and introduction (section 3) and conclude with a discussion of the prospects, discussions and open questions (section 4). 


\section{Self-interacting Scalar in AdS}
In this paper we will propagate a probe scalar field with the action
\begin{eqnarray}
S=\int d^4x \sqrt{-g} \left(\frac{1}{2} \nabla_\mu \phi \nabla^\mu \phi + V (\phi 
) \right)
\end{eqnarray}
on a non-dynamical AdS
, with reflecting boundary conditions. The dynamics of this system has previously been studied in \cite{Basu}, as a holographic model for quantum quench. We will take the potential to be that of a massless scalar $m=0$ in what follows, but with a quartic coupling
\bea
V(\phi)= \frac{\lambda}{4!} \phi^4.
\eea
We work with global AdS in the coordinates: 
\begin{eqnarray}
ds^2= \sec^2 x (-dt^2+dx^2+\sin^2 x d\Omega^2).
\end{eqnarray}
Specifically, we look for spherically symmetric solutions of the non-linear wave equation arising from the Lagrangian. The non-trivial (radial) part of the equation reduces to
\begin{eqnarray}
\phi^{(2,0)}
+\Box_s \phi 
\equiv \phi^{(2,0)}
-\phi^{(0,2)}
-\frac{2}{\sin x \cos x}\phi^{(0,1)}
=-\frac{\lambda}{6 \cos^2(x)} \phi 
^3 
\end{eqnarray} 
We will find it useful to have solutions of the eigenvalue equation $\Box_s e_j =\omega_j^2 e_j$  for the spatial Laplacian $\Box_s$:
\bea
e_j(x)=4 \sqrt{\frac{(j+1) (j+2)}{\pi }} \cos ^3(x) \, _2F_1\left(-j,j+3;\frac{3}{2};\sin ^2(x)\right), \\ \omega_j^2=(2j+3)^2. \hspace{1.5in}
\eea

We will evolve the wave equation starting with various initial conditions, and we will be interested in keeping track of the total energy as well as the energy per mode. To avoid overall energy non-conservation and losses due to numerical artifacts, we will use a symplectic Hamiltonian formalism for dealing with the equations of motion numerically. The Hamiltonian density 
takes the form
\bea
{\cal H}=4 \pi  l^2 \tan ^2x \left(\frac{l^2 V(\phi )}{\cos ^2(x)}+\frac{1}{2} \phi ^{(0,1)}(t,x)^2\right)+\frac{\Pi (t,x)^2}{8 \pi  l^2 \tan ^2(x)},
\eea
where we have temporarily re-instated the AdS scale $l$ for clarity, and
$\Pi$ is the canonical momentum. We have done the integration over the angles of the sphere in AdS already: in other words, this Hamiltonian density when integrated over the $x$ gives us the full Hamiltonian of the system.
In terms of the mode sum, this takes the form
\begin{equation}
H=\frac{1}{2} \sum_{i=0}^{\infty} \big(\dot A_i(t)^2+\omega_i^2 A_i(t)^2\big)+\sum_{i,j,k,l=0}^{\infty} {\cal C}_{ijkl} A_i(t) A_j(t) A_k(t)A_l(t) \label{Ham}
\end{equation}
where the ${\cal C}_{ijkl}$ coefficients can be straightforwardly computed.

\section{Holographic Strong Stochasticity Threshold}\label{sstsec}

\subsection{Multi-scale Analysis}

At low energy (i.e. $A_0 \ll 1$) resonant transfer between modes is the most important mechanism of energy transfer \cite{Chirikov}. The naive perturbation theory in $\lambda$ breaks down at large-$t$ due to resonances. However in often-used multi-time analysis, one separates the slow and fast parts of the degrees of freedom by defining $A_i(t) \approx B_i( \tau) e^{i \omega t}+\bar{ B_i(\tau)}e^{-i \omega t} $, where $\tau \equiv \epsilon^2 t$ is the slow-moving time. If we take a $t$-average of the Lagrangian over a fundamental period and assume $\tau$ dependent terms to be constant. We get,
\begin{align}
L_{MS}=i \sum \omega_i (B_i \dot{\bar{B_i}}-B_i \dot{\bar{B_i}})+\sum  {\cal C}_{ijkl} \bar{B_i(\tau)} \bar{B_j(\tau)} B_k(\tau) B_l(\tau)\label{ttf-lag} ,
\end{align}
with $\omega_{i}+\omega_{j}-\omega_{k}-\omega_{l}=0$ in the second term. Here, in the interaction part only $t$-independent terms satisfying the resonance condition survive.
One can systematically include corrections by introducing other time scales in (\ref{ttf-lag}), but even the zero'th order theory gives us information about how energy is transferred across resonances. 

One interesting observation is that other than the usual time translation invariance, (\ref{ttf-lag}) has a set of enhanced symmetries. From the dilatation symmetry of the form $B_k(\tau) \rightarrow \alpha B_k(\frac{1}{\alpha^2} \tau)$, one may naively presume that the thermalization time goes inversely with the square of initial amplitude. However (\ref{ttf-lag}) has some other enhanced symmetries that are not possessed by the full system, so it seems possible that these other symmetries might forbid the thermalization of the theory in the multi-scale approximation.
Here we list the conserved charges and the corresponding symmetries:
\bea
Q_0=\sum B_k \bar{B}_k, \quad {\rm Symmetry:\ } B_k \rightarrow e^{i\theta} B_k , \hspace{1.5in} \\
Q_1=\sum k B_k \bar{B}_k,\quad {\rm Symmetry:\ } B_k \rightarrow e^{i k \theta} B_k ,\hspace{1.4in} \\
E=\sum_{\omega_i+\omega_j-\omega_k-\omega_l=0}  {\cal C}_{ijkl} \bar{B_i(\tau)} \bar{B_j(\tau)} B_k(\tau) B_l(\tau), \quad {\rm Symmetry:\ } t \rightarrow t+\alpha.\hspace{0.5in}
\eea

A closely related fact is that the multi-scale theory has a set of quasi-periodic solutions given by $B_k(t)=\alpha_k \exp(-i(\beta_0 t+(i-1)\beta_1 t))$. Plugging this into the multi-scale equation of motion we can solve for $\beta_0,\beta_1,\alpha_k$, as also observed in \cite{Buchel}. The enhanced symmetries and the accompanying slow thermalization are consistent with the previous observation \cite{Pettini2} that the thermalization times are of the order of $e^{-1/A^{\delta}}$ in some statistical systems, where $\delta$ is some positive constant. 

\begin{figure}[h]
\subfigure[The upper envelope of $\dot \phi(x=0)^2$, is qualitatively identical to figure 3 in \cite{Buchel}.  \label{ttf3} The fine structure here in the envelope is due to a finer smoothing scheme.]{\includegraphics[width=0.5\textwidth]{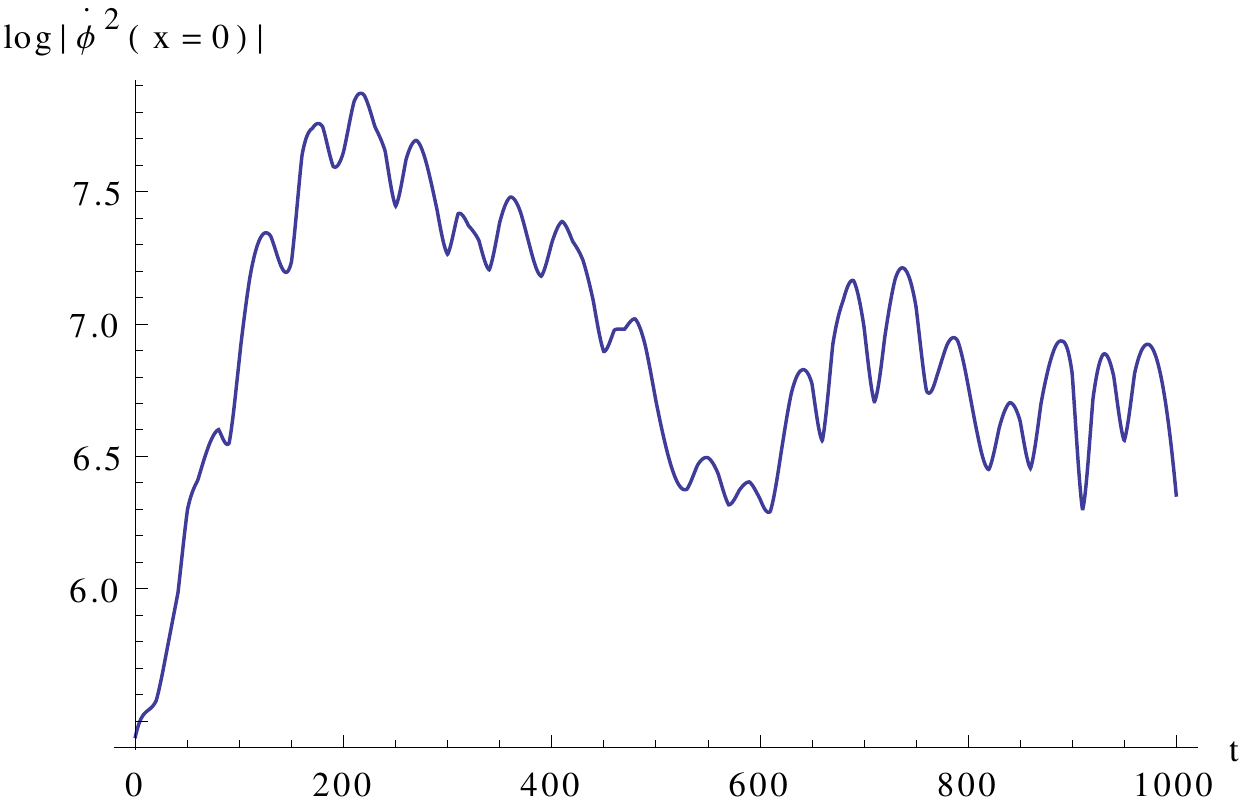}} \ \
\subfigure[Running time-average energy per mode $\bar E_J$ vs $t$ 
for equal-energy two-mode initial data. \label{ttf5} This is qualitatively identical to figure 5 in \cite {Buchel}. The curves correspond to $j=0,1,2,3$ top to bottom.]{\includegraphics[width=0.5\textwidth]{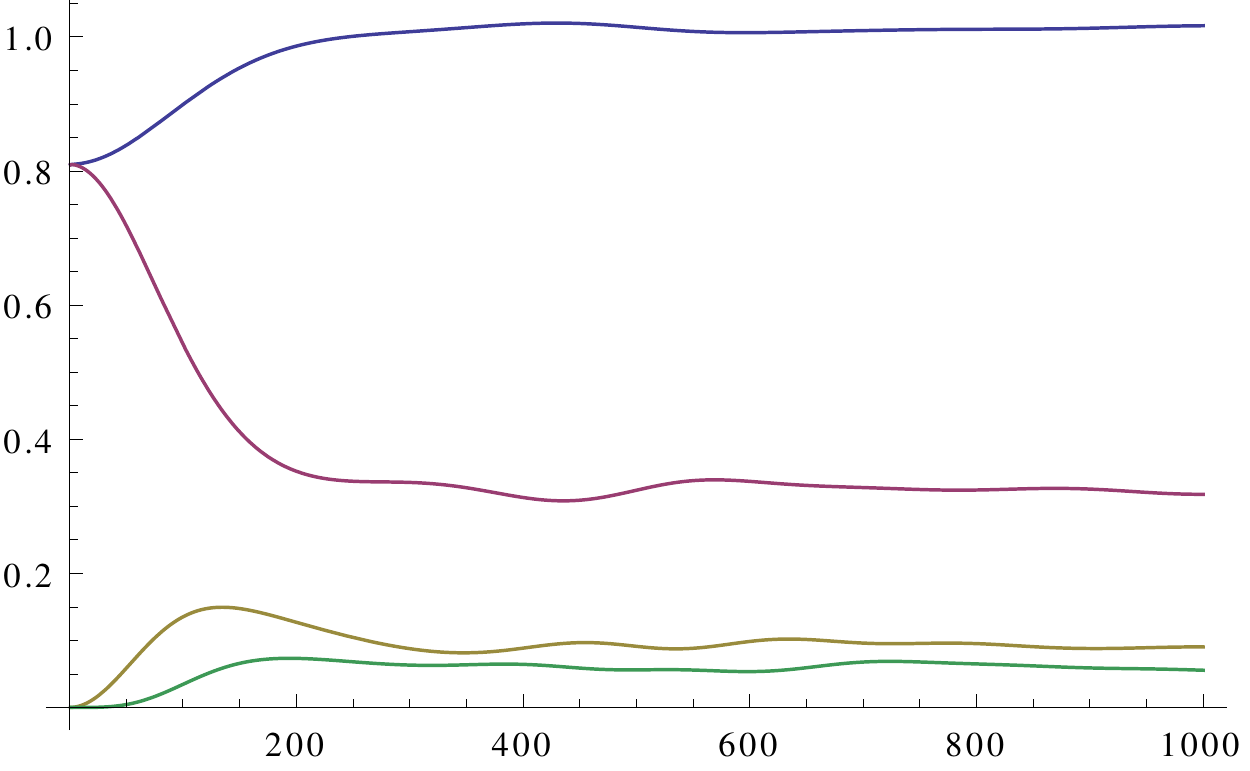}} 
\caption{Both these plots have been made with initial data that contained equal energy in the $j=0, 1$ modes.}
\end{figure}

To provide evidence that our truncated system can capture the relevant features of the gravity system, we first note that the scalar evolution is quasi-periodic for a long time, as already emphasized. This was the case in \cite{Buchel} as well. We also present plots of some of the quantities. These have qualitative similarities with the results in \cite{Buchel}. In particular our figures \ref{ttf3}, \ref{ttf5} (which were made using a Two-Time Formalism), should be compared to figures 3 and 5 in \cite{Buchel}.

\subsection{Truncated Model}

As a preliminary exercise, we will first study a system where we truncate the full system and restrict the modes to only the first $9$ energy eigenstates: ie., the sum in  (\ref{Ham}) will be from 0 to 8\footnote{We also study the case where we restrict to $17$ modes, but the results are qualitatively similar.}. We integrate the resulting Hamilton's equations of motion with an initial condition $A_1(0)=A_0$ while setting the rest of the initial  $A_i$'s and all the conjugate momenta to zero. To do this, without incurring fictitious losses in the total energy due to numerics, we resort to a sympletic integration of the equations that preserves the phase space volume.

Due to the presence of non-linearity, the dynamics of the system is not exactly integrable. 
However, this fact does not 
tell us much about the approach of the system towards thermalization as it evolves. In fact, thermalization is a somewhat tricky idea to precisely capture, so we will use the following approach to define it. A thermal system shows equipartition of energy among modes, i.e. the virial theorem holds. For the kinetic part it means that the time average 
\bea
\langle  \dot A_i(t)^2  \rangle_t \sim 1/t \int_{0}^{t} \dot A_i(\tau)^2 d\tau
\eea 
for large-$t$ is independent of the mode number. So we will use this time-avegared kinetic energy per mode as a quantity that keeps track of the approach to thermalization. 

\begin{figure}[h]
\subfigure[$A_0(0)=0.7$]{\includegraphics[width=0.5\textwidth]{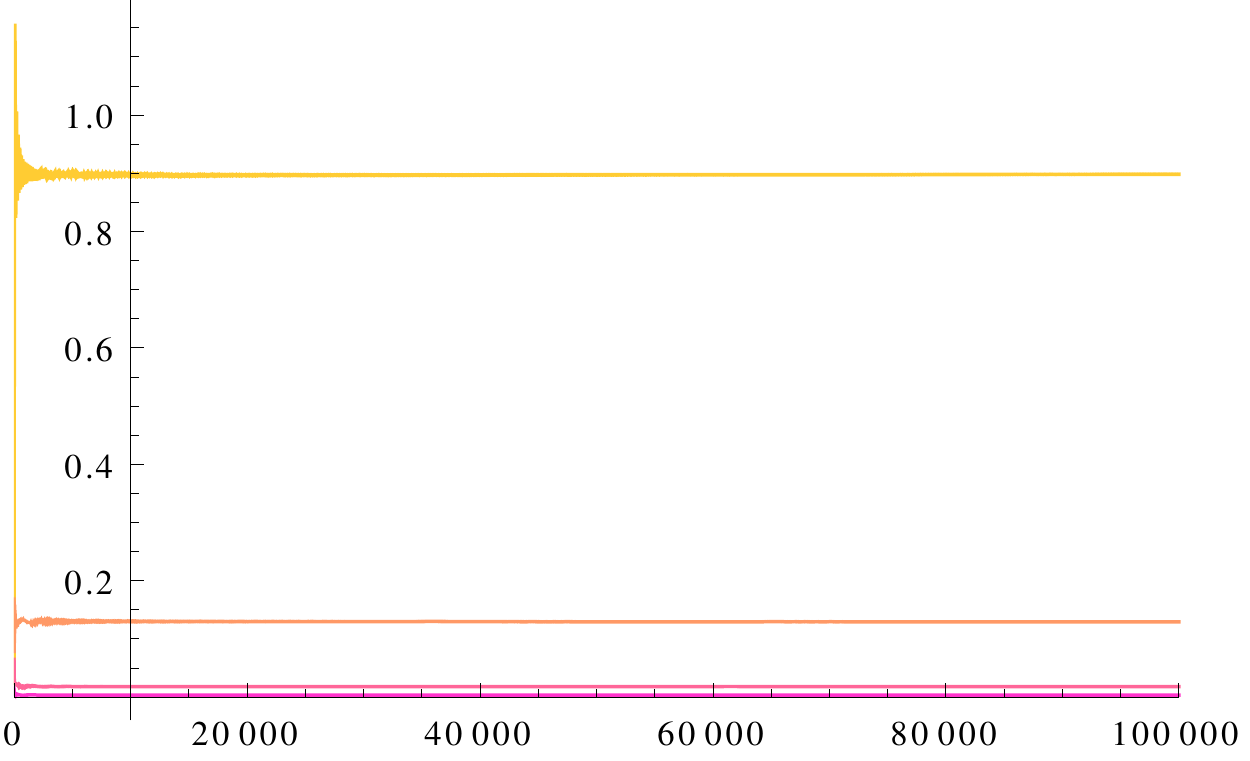}} \label{truncnotherm}
\subfigure[$A_0(0)=0.71$]{\includegraphics[width=0.5\textwidth]{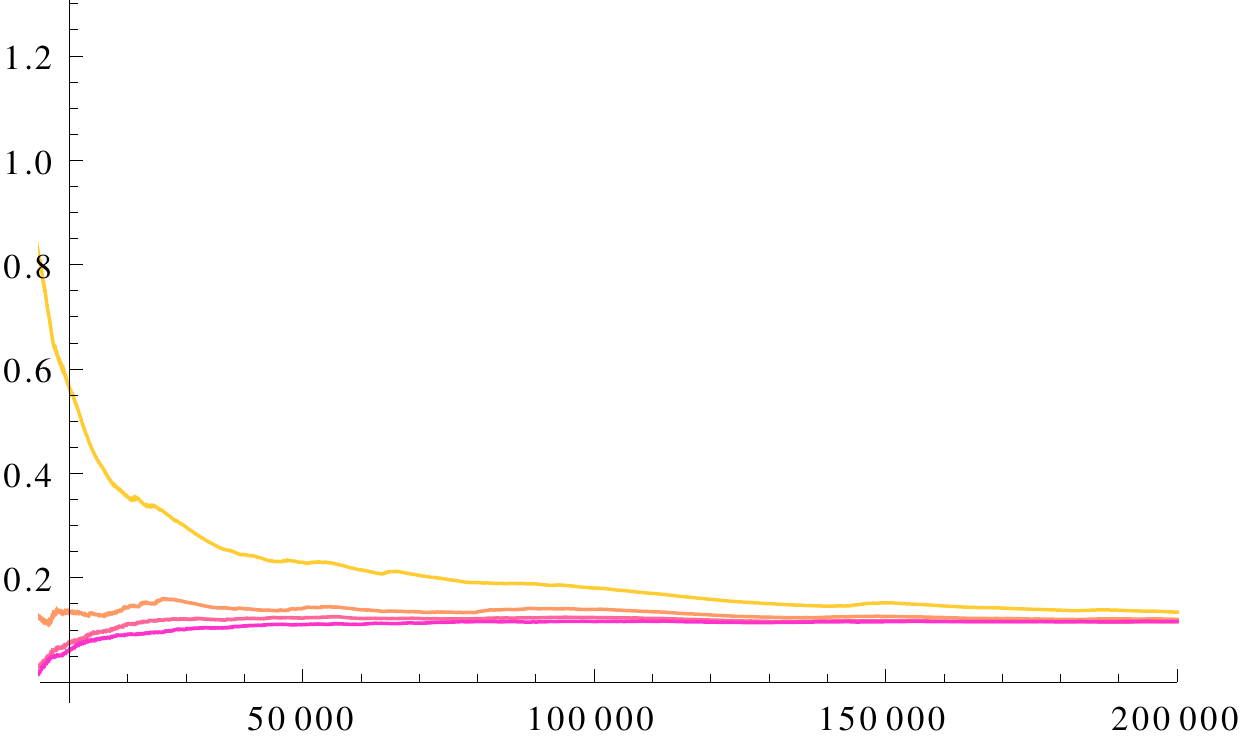}} \label{trunctherm}
\caption{Plots of time average of kinetic energies (in the truncated model) for modes- $0,1,2,3$ from the top.}
\end{figure}

\begin{figure}[h]
\centering
\includegraphics[width=0.4\textwidth]{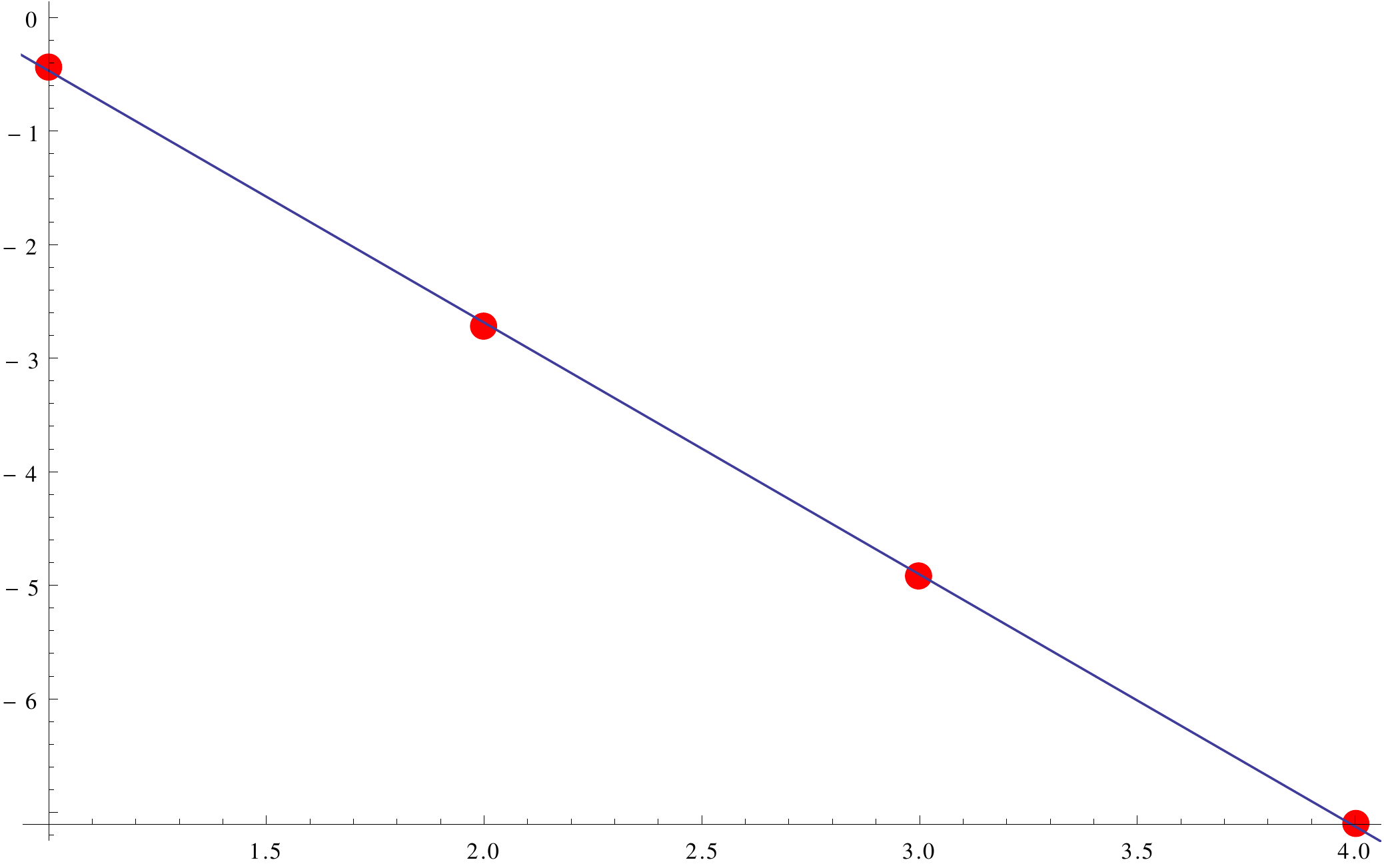}
\caption{Mode number vs lograthim of $\langle\dot A_n(t)^2  \rangle_t$ for large $t$.}
\label{fig:plt0}
\end{figure}

We have investigated the full time evolution of the truncated theory. When the initial amplitude is small, within the scope of our numerics we do not see any tendency of the theory to equilibrate even after long times of integration. In fact, we always find that the average kinetic energy of high enough modes are exponentially suppressed, i.e. $  \log (\langle\dot A_n(t)^2  \rangle_t) \sim -O(n)$ for large-$n$ (see Fig \ref{fig:plt0}). 

On the other hand, according to the theory of \cite{Chirikov} (see also \cite{Pettini}) one expects that at higher energy direct energy transfer between modes (outside resonant couplings) would be important. For example in our case, we find that for mode $j=0$,  if we start with an initial value of $A_0 > 0.707$(approx), the theory quickly thermalizes. We have considered various other modes and combinations of modes as well: a similar story (but with other values of the threshold amplitude/energy) holds also in these cases. This is evidence for a strong stochasticity threshold (SST). 

In the mode-truncated system that we discussed in this subsection, the SST is a very sharp feature and the transition is quite drastic (almost as if it were a phase transition). When we keep all the modes and do not do the truncation, we will again see that there is a big qualitative difference in thermalization times as we cross the threshold energy, but the transition is less sharp and more reminiscent of a crossover rather than a phase transition. This is what we discuss in the next subsection.


 
\subsection{Full theory}

Here we look at the time evolution of full scalar equations in $AdS$ numerically.  To understand the time evolution numerically, we discretize the hamiltonian on a lattice with $N=256$ ( and also $512$ points) and time evolve the Hamilton's EOM's using a simplectic Runge-Kutta algorithm. 

We start with various initial conditions by turning on various combination of modes. We will confine ourselves to low lying modes and highest mode we would go up to is ten. Keeping the ratio of amplitudes of modes fixed, we change the overall amplitude. We always see a SST, below which the theory does not have effective thermalization and over which the theory quickly thermalizes. 

\begin{figure}
\subfigure[$A_8(0)=1.04$]{\includegraphics[width=0.5\textwidth]{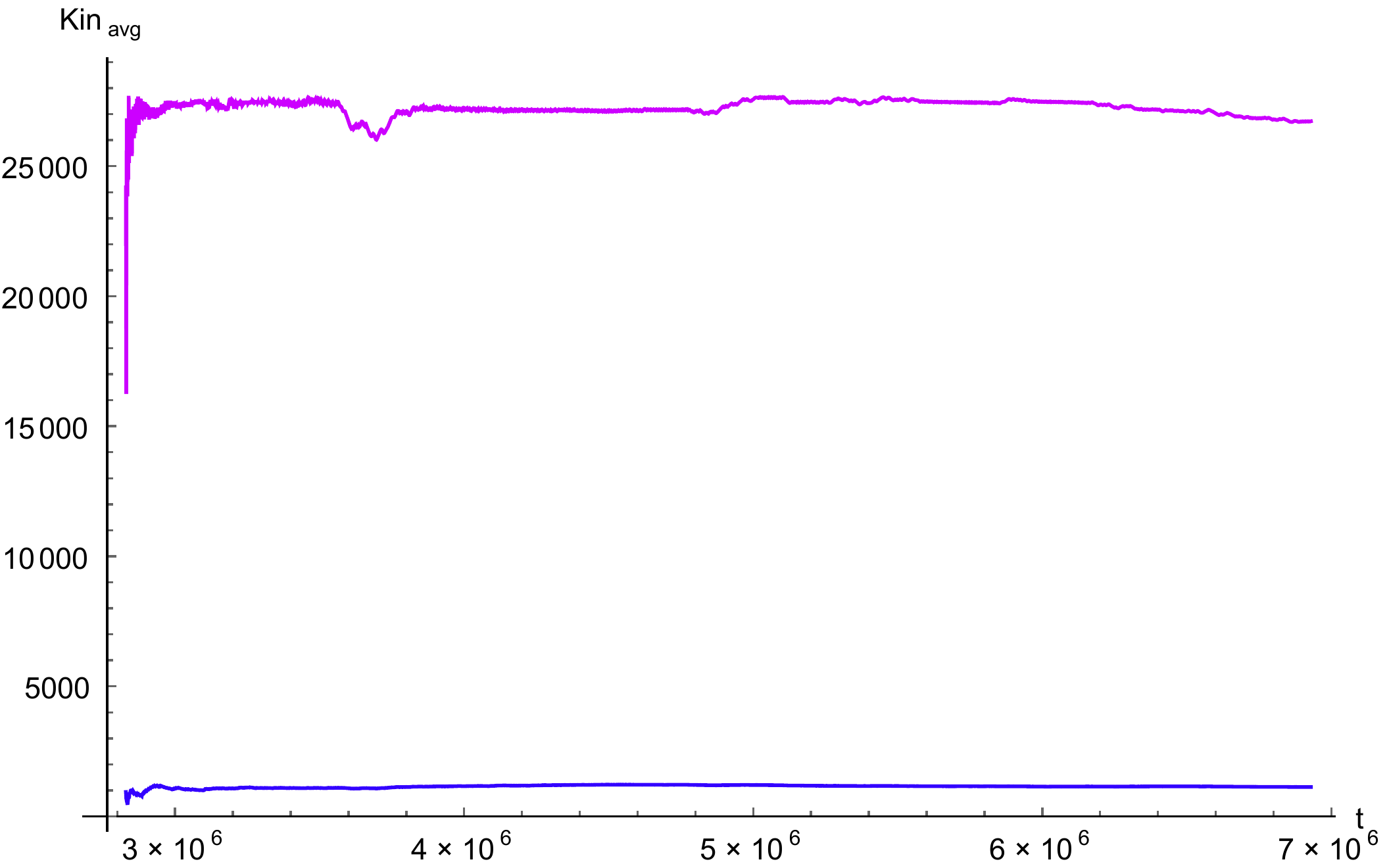}\label{fg104}}
\subfigure[$A_8(0)=1.08$]{\includegraphics[width=0.5\textwidth]{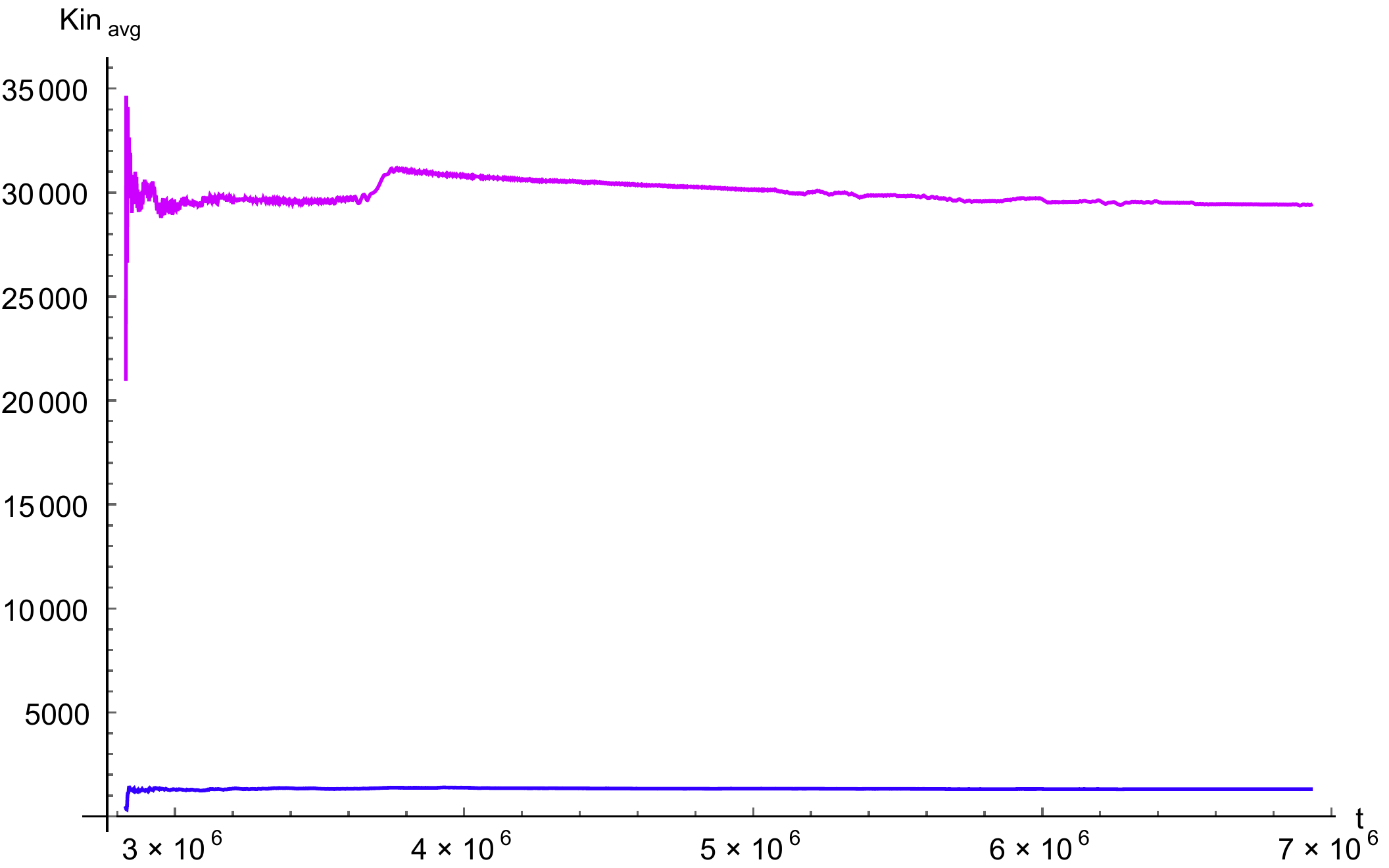}\label{fg108}}
\subfigure[$A_8(0)=1.12$]{\includegraphics[width=0.5\textwidth]{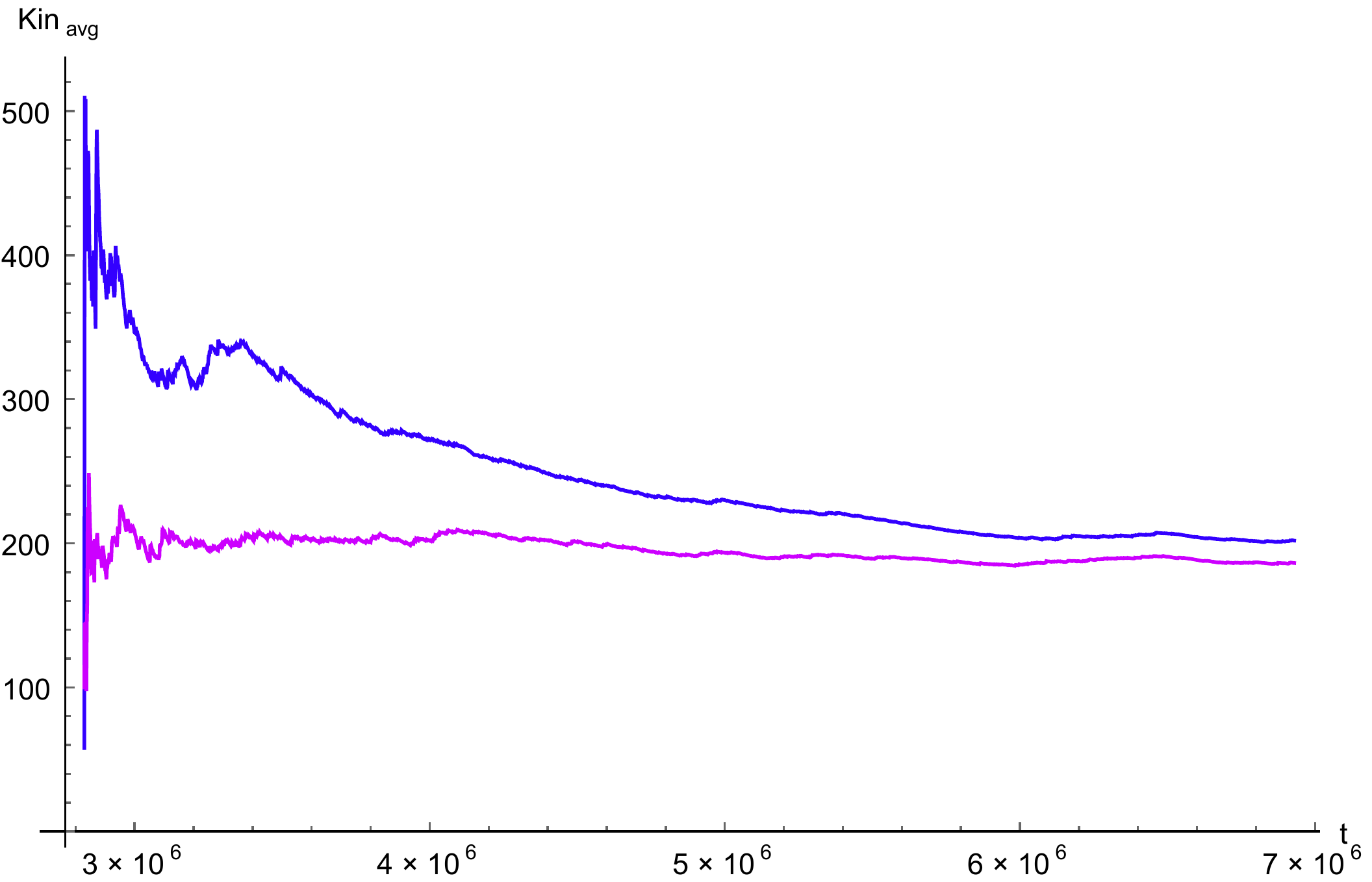}\label{fg112}}
\subfigure[$A_8(0)=1.16$]{\includegraphics[width=0.5\textwidth]{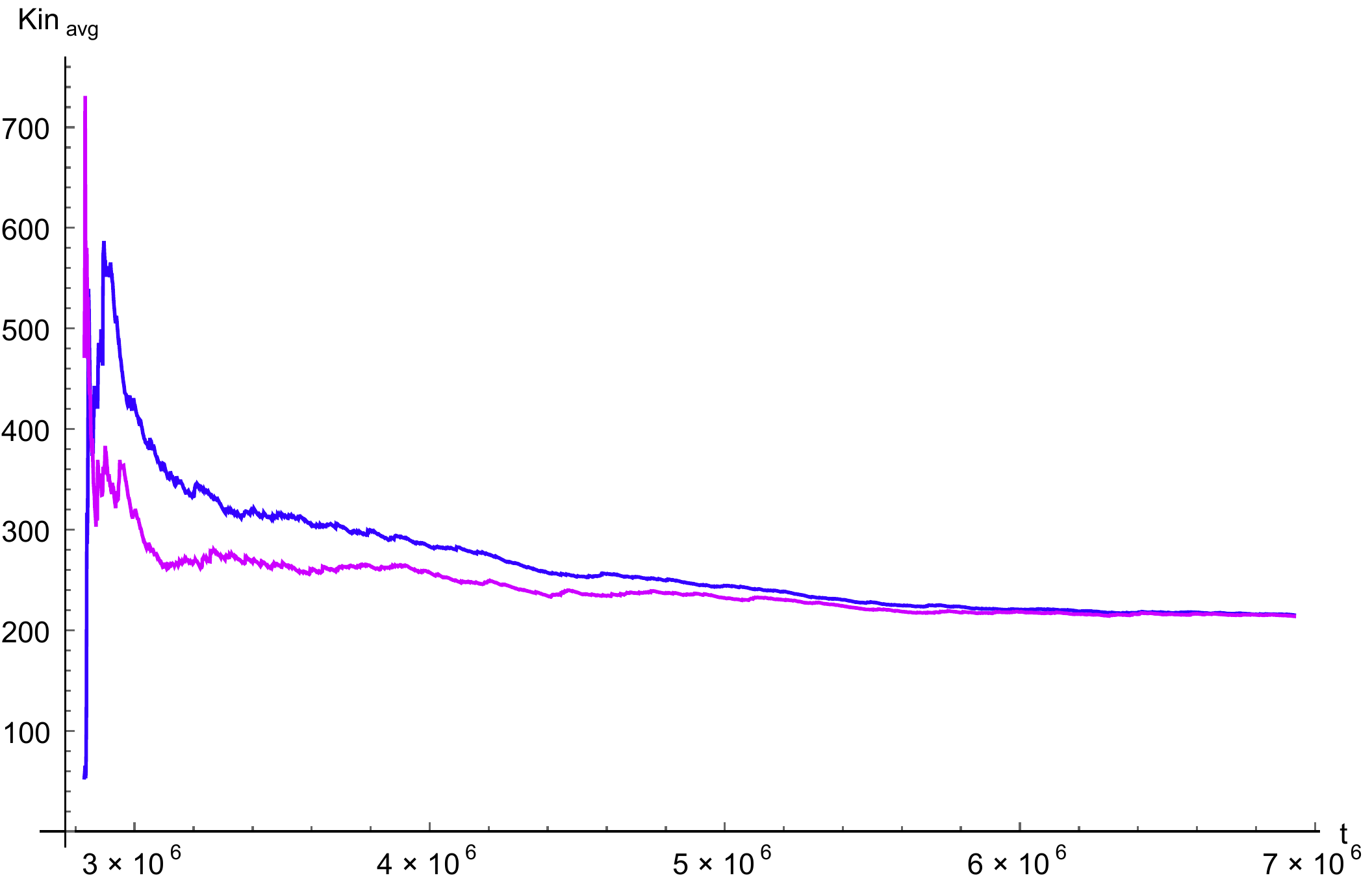}\label{fg116}}
\caption{Plots of time averages of kinetic energy with time, for modes- $6$(top) and $7$(bottom).}
\end{figure}

We here describe the main story with only the eighth mode 
turned on initially. We look at the time evolution of the system for a time scale of the order $t=1\times10^{8}$ (to compare $\omega_0=3$ and $T_0=\frac{2\pi}{\omega_0}\approx 2.094 $). Technically thermalization time may be tricky to define. Here we would not go into the technicalities. For our purpose we will pick two modes (taken to be $7,8$ in the plots) and look at their average kinetic energy (averaging started at $t\approx 2.75\times 10^6$ ) to have a better idea of thermalization. For a small initial amplitude the theory does not show a tendency to thermalize within the integration time scale Fig \ref{fg104}. The situation changes after quickly after $A_8(0) \approx 1.08$ (Figure-\ref{fg112}). Between Fig-\ref{fg104} and Fig-\ref{fg116}, we clearly see a crossover behavior and a few percent increase of initial energy leads to a huge decrease in thermalization time.

\begin{figure}
\subfigure[$A_8(0)=1.08$]{\includegraphics[width=0.5\textwidth]{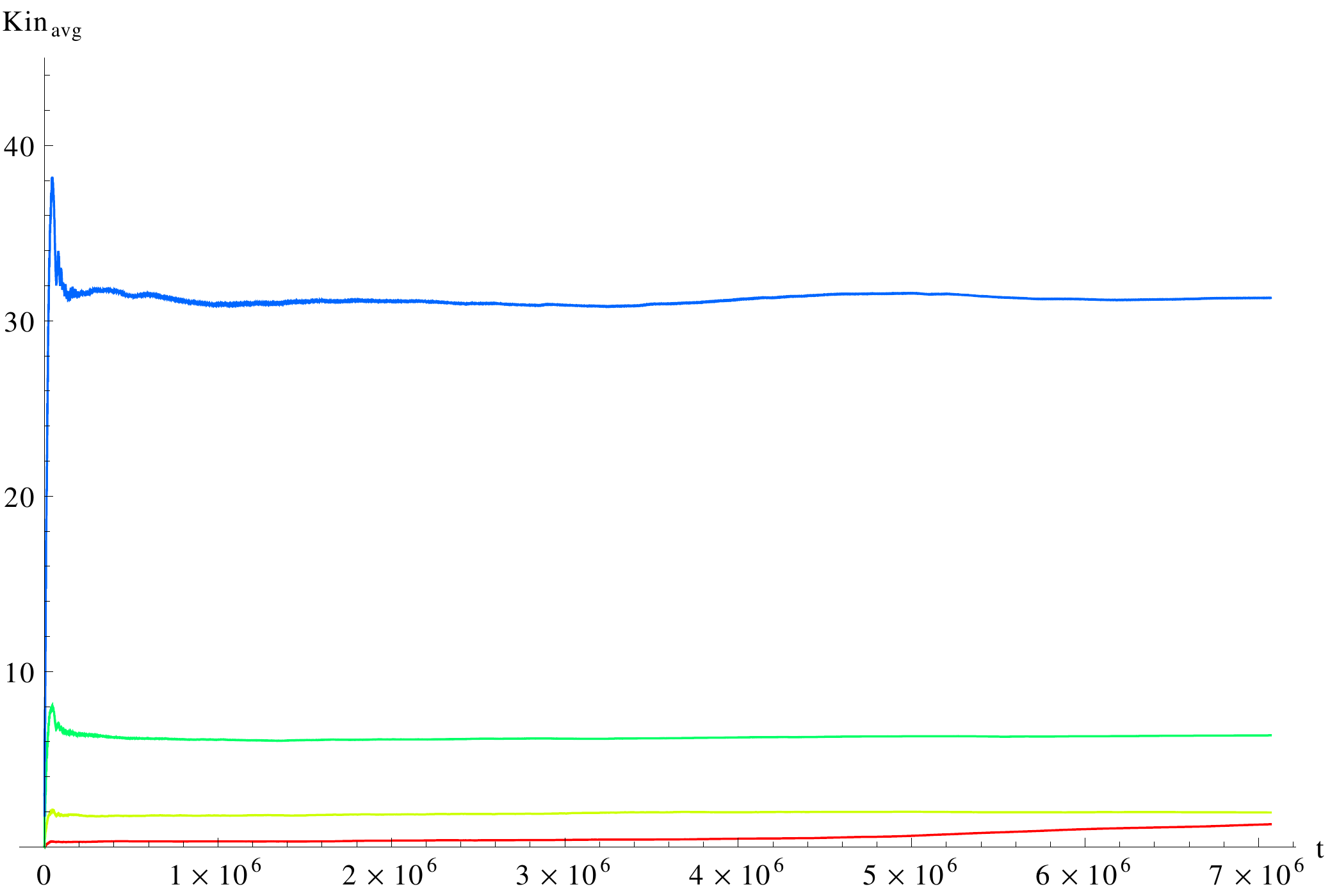} \label{F1.08}}
\subfigure[$A_8(0)=1.16$]{\includegraphics[width=0.5\textwidth]{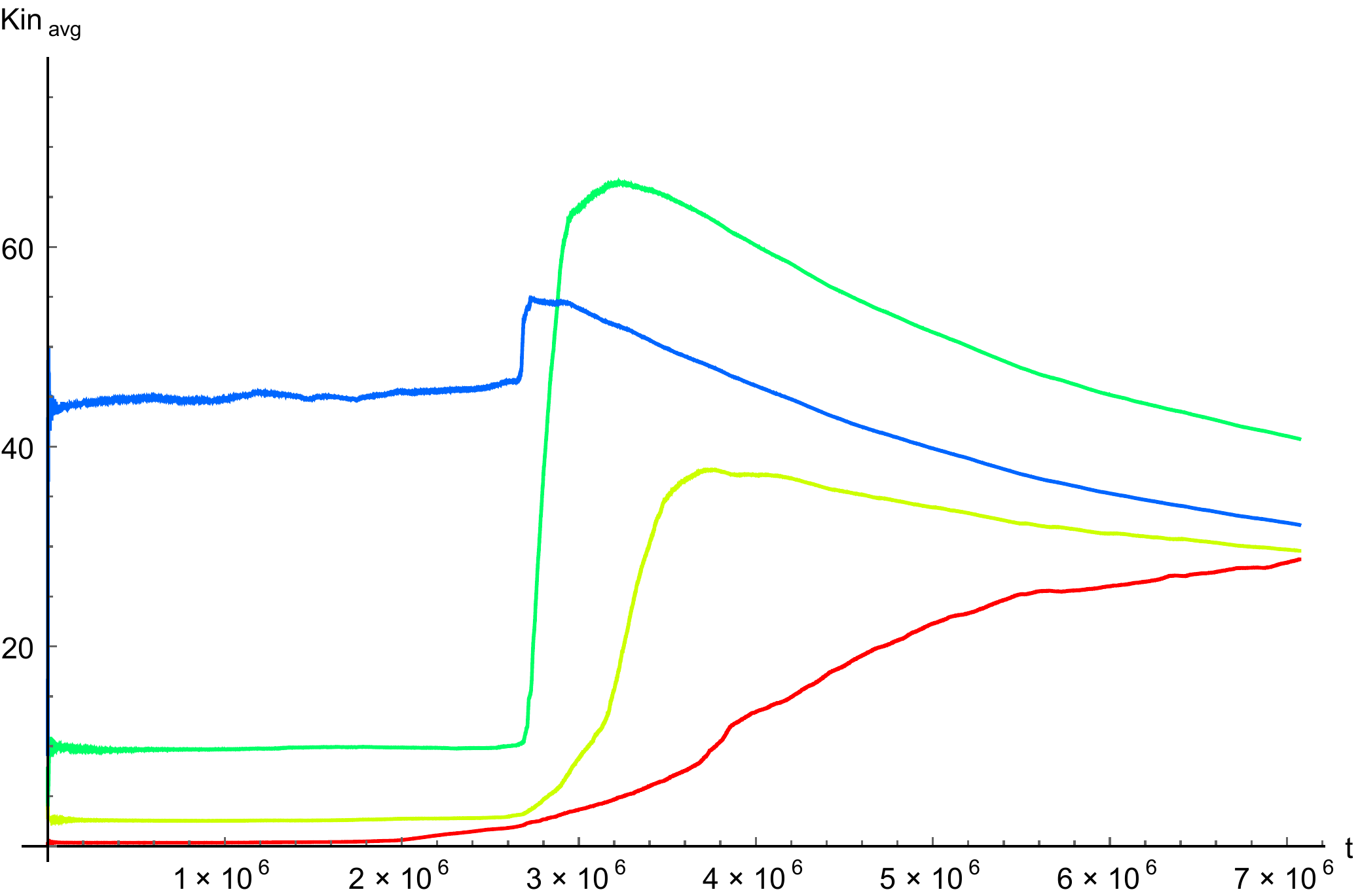} \label{F1.16}}
\subfigure[$A_8(0)=1.28$]{\includegraphics[width=0.5\textwidth]{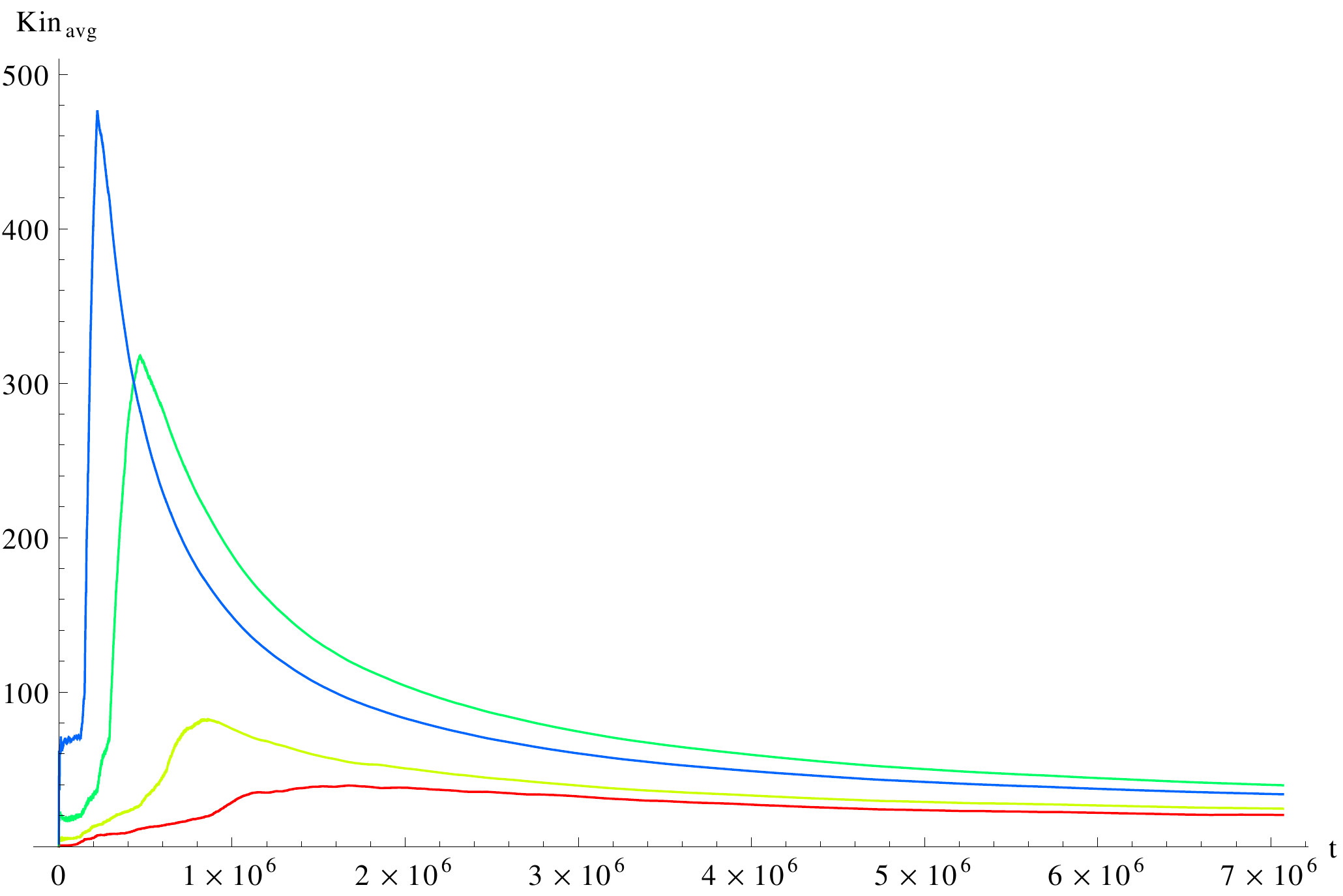} \label{F1.28}}
\subfigure[$A_8(0)=1.5$]{\includegraphics[width=0.5\textwidth]{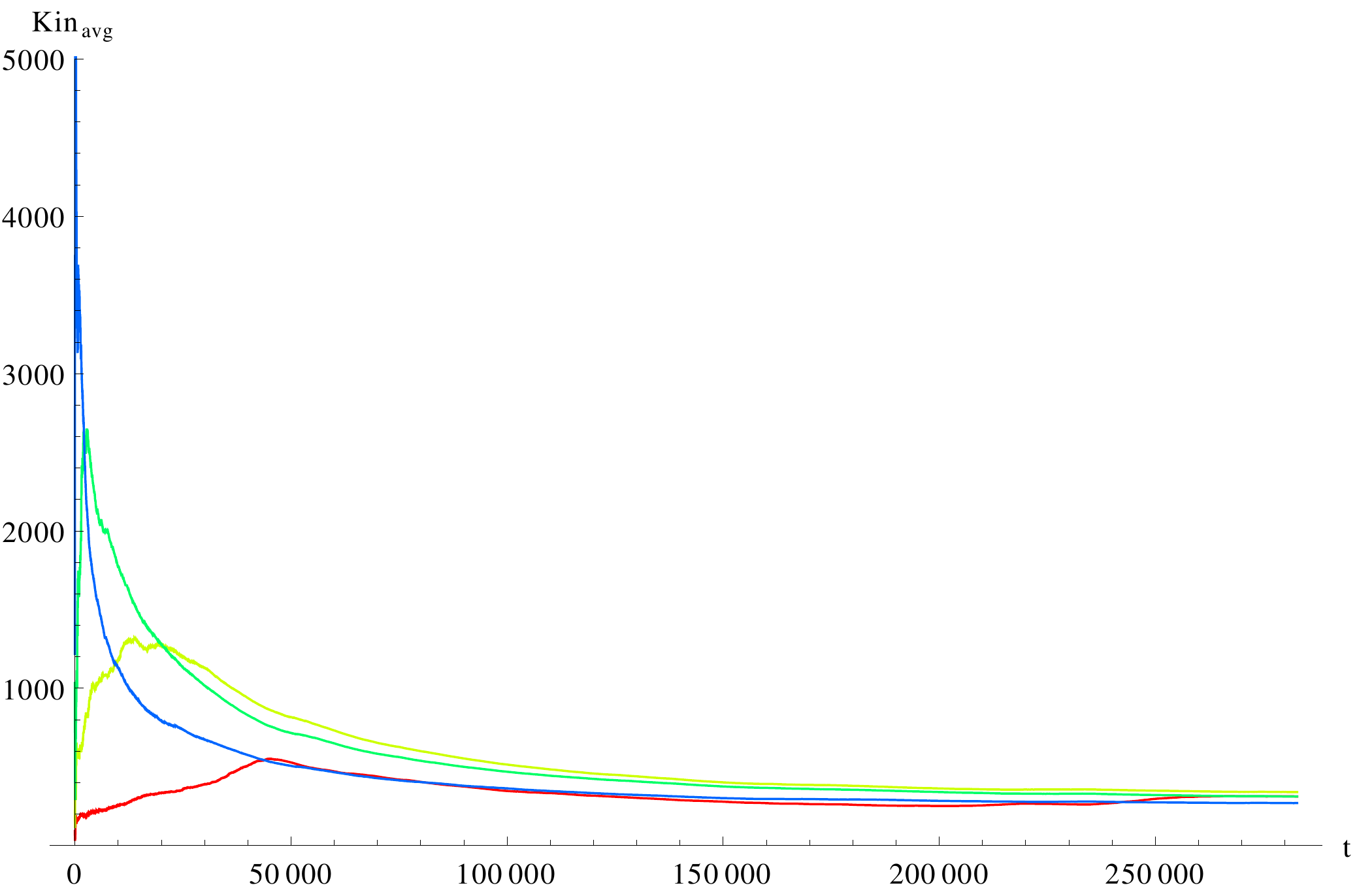} \label{F1.5}}
\caption{Plots of time average kinetic energy with time, for the modes- 0, 2, 4, and 6 (from the top at left).}
\end{figure}

It is also interesting to note how thermalization happens. For that we made detailed plots of average kinetic energy of many modes. For low enough initial amplitude, initially for some time the system is trapped in a non-thermal steady state. For example, in Fig-\ref{F1.16}, before the average kinetic energies thermilizes, they level out like in Fig-\ref{F1.08}. A vestige of this phenomenon can also be seen in Fig-\ref{F1.28}. One possible explanation is that system was trapped in valley in the phase space with a narrow escape, and when the phase space point reaches that escape location, the kinetic energy shoots up. This phenomenon is not observed for $A_8(0)>1.3$ (see Fig- \ref{F1.5}). Also thermalization behavior of the system is only weakly dependent on the initial energy after the thermalization threshold. It is believed that the dominant channel of energy transfer between modes changes from resonance transfers to non-resonant transfer after the thermalization threshold \cite{Pettini}. 

Another noteworthy feature of the curves is the upward rise in the time-averaged kinetic energy per mode close to the onset of thermalization. We do not believe that this is a numerical artifact\footnote{The timescale of the rise is a fraction of the total time-range, and is very many orders of magnitude bigger than the integration time-step. Besides, we have checked that the total energy in the system in our sympletic integrator is constant to within negligible numerical errors.}, but we also do not understand the physical significance of this. But it is tempting to speculate that this is due to an escape from a trapped region in phase space as we mentioned in the previous paragraph.

\begin{figure}
\subfigure[$A_8(0)=1.04$]{\includegraphics[width=0.5\textwidth]{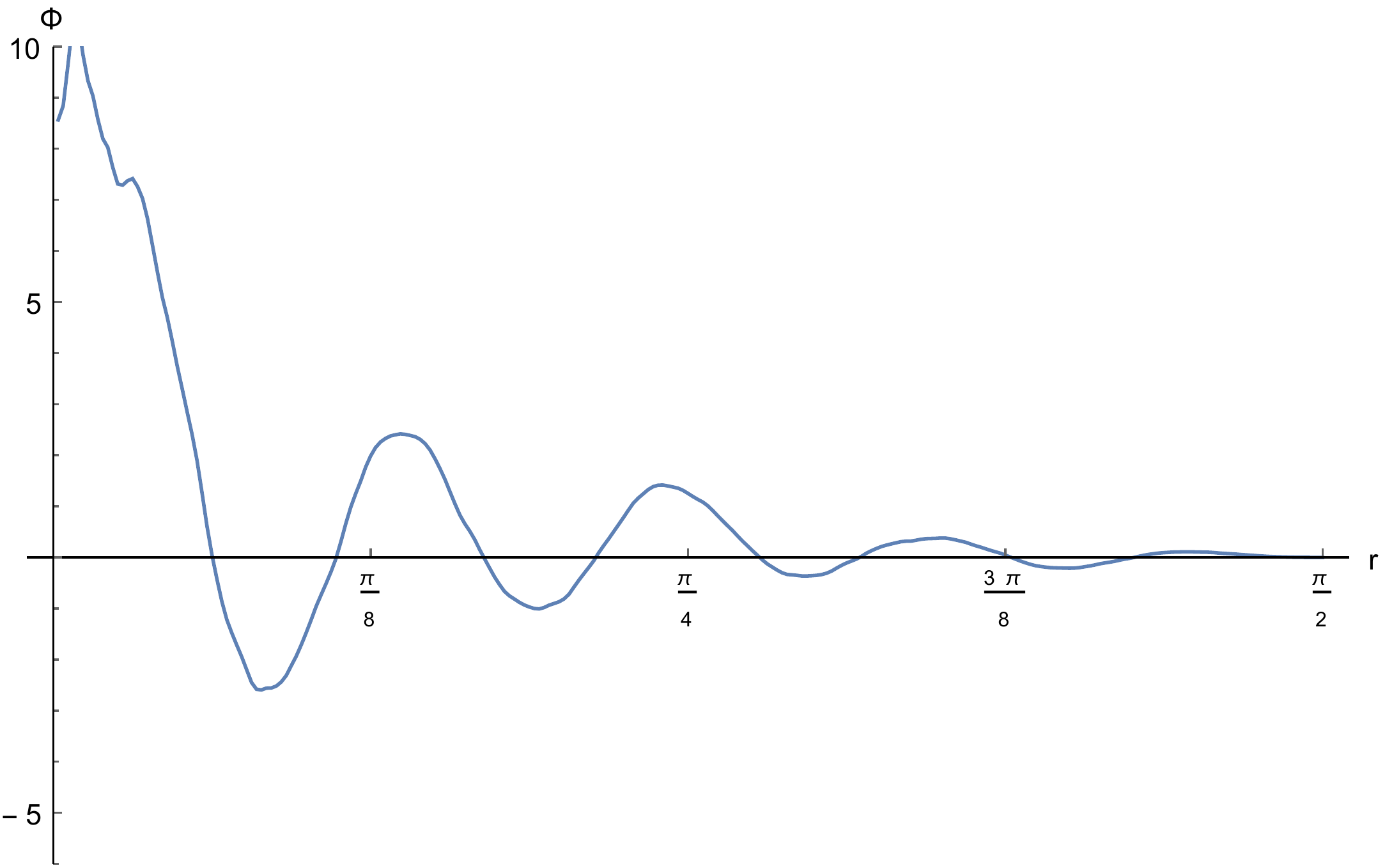}} \label{FS1.04}
\subfigure[$A_8(0)=1.08$]{\includegraphics[width=0.5\textwidth]{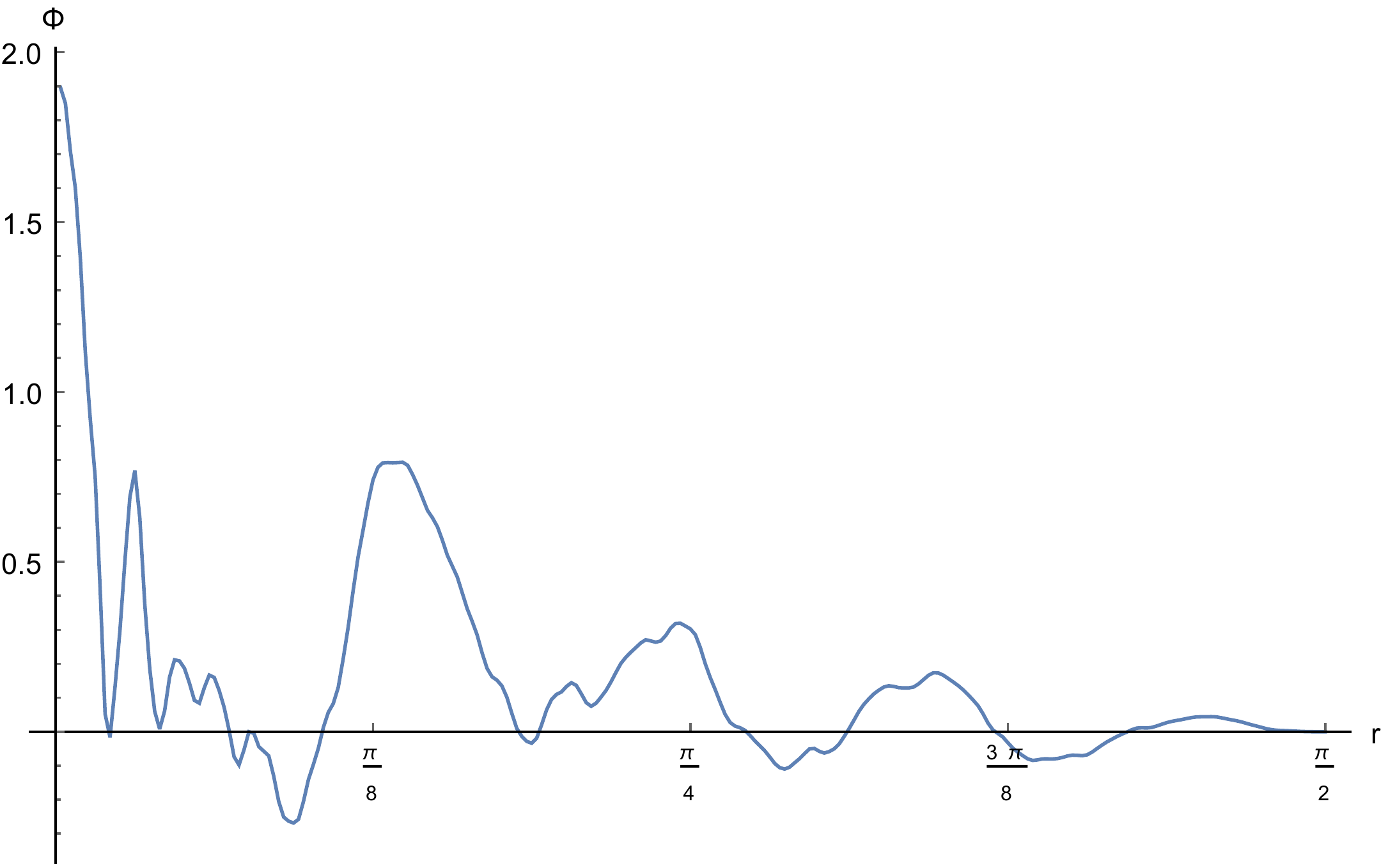}} \label{FS1.08}
\subfigure[$A_8(0)=1.12$]{\includegraphics[width=0.5\textwidth]{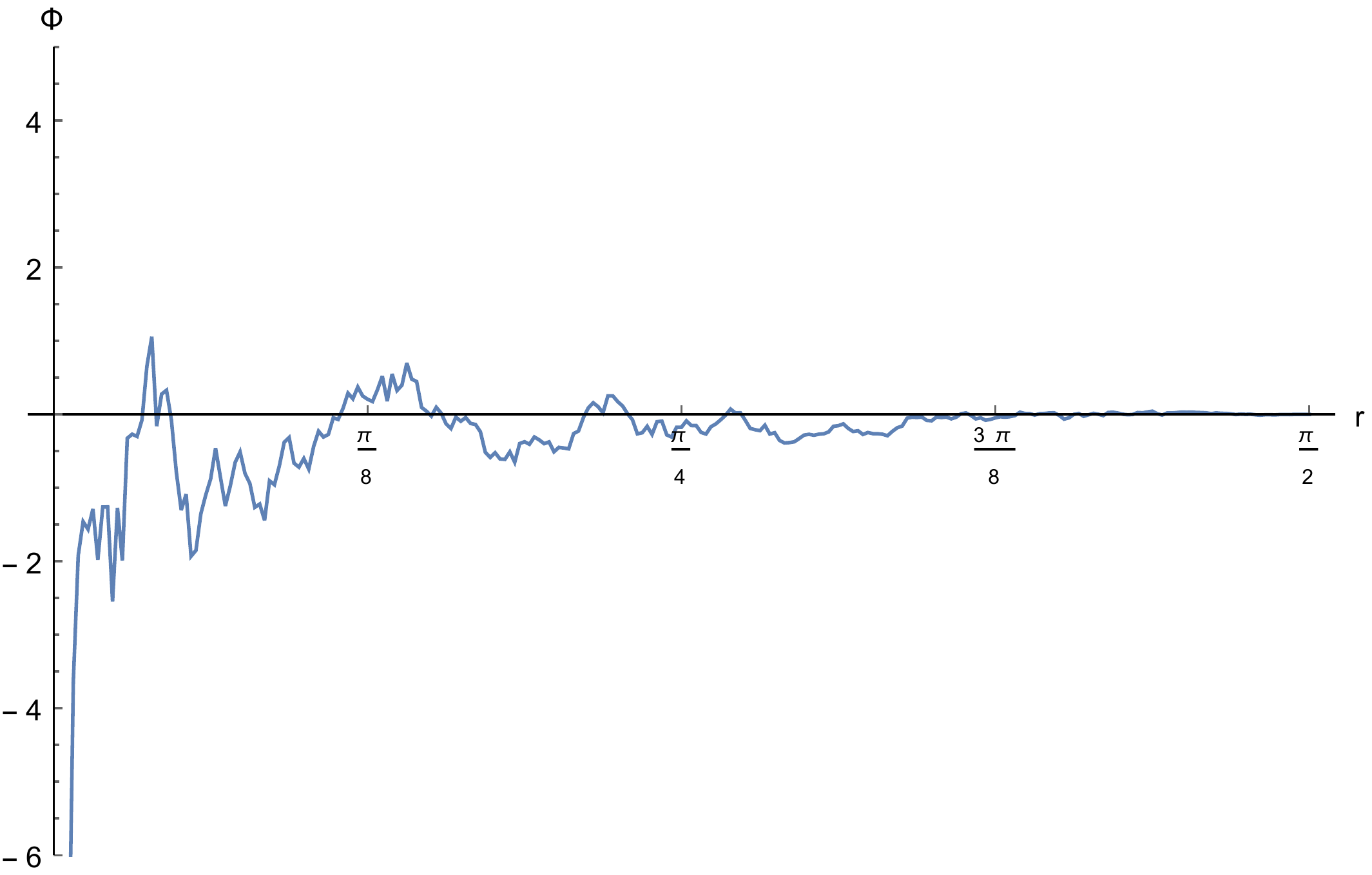}} \label{FS1.12}
\subfigure[$A_8(0)=1.16$]{\includegraphics[width=0.5\textwidth]{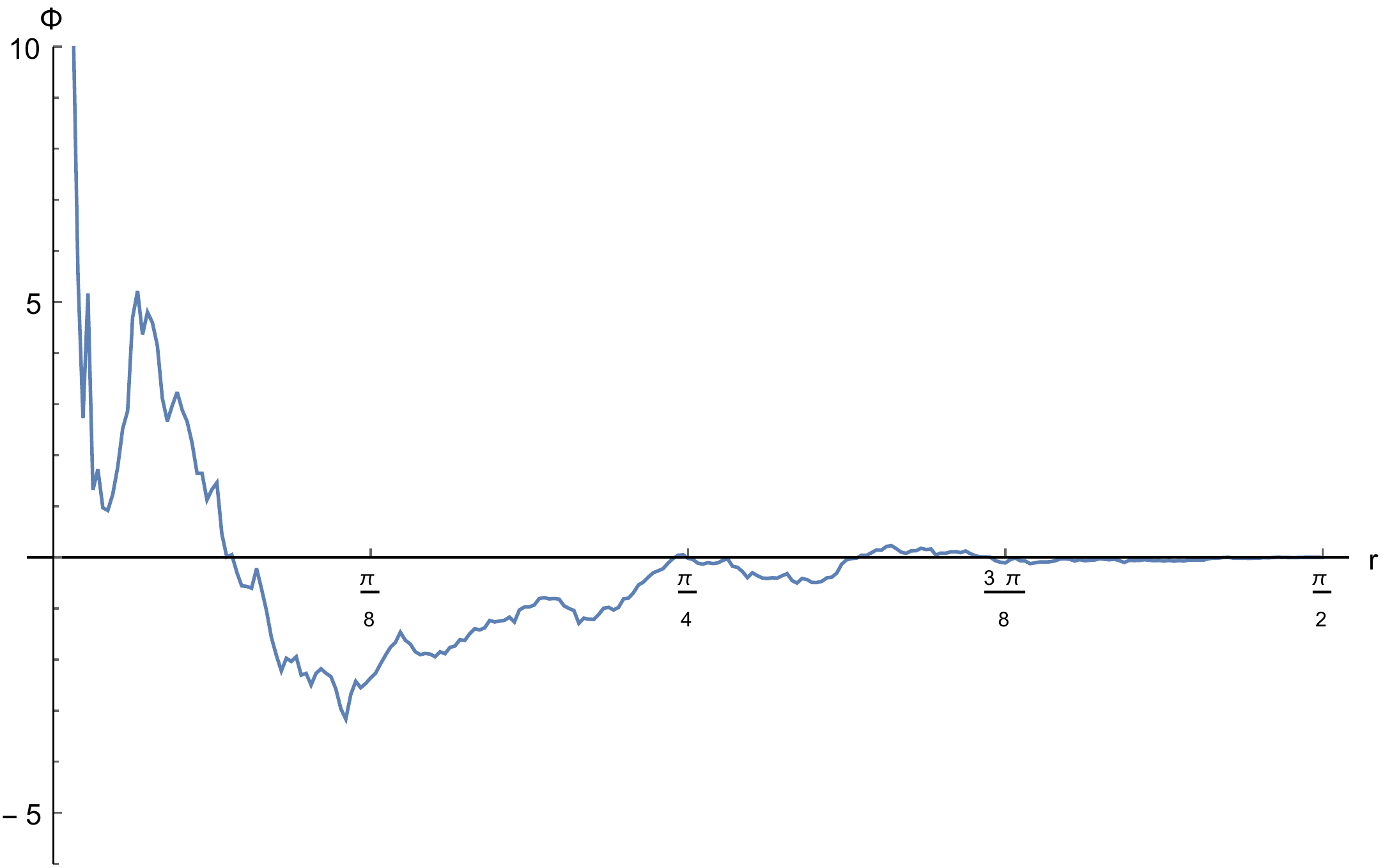}} \label{FS1.16}
\caption{Radial profile of the scalar field at $t\approx 6.283 \times 10^5$.}
\label{FSall}
\end{figure}

To capture some of the features of the threshold, we also show the spatial profiles of the scalar field after a large number of time steps, above and below the SST: the drastic transition of the waveform from quasi-periodic to chaotic is evident (see Fig \ref{FSall}), as expected of a system that is close to thermalization. It is also noteworthy that unlike in the case of the mode-truncated system, after thermalization, the system relaxes to zero average energy per mode as it should. We have checked in specific cases that the average energy per mode comes down to very low values { (1\% of original)}.  But the computational time required to evolve the system for long times is large, so we have not checked it for all cases. 
  
\section{Discussion}


In this paper, we have demonstrated the existance of a Holographic Strong Styochasticity Threshold for thermalization of an interacting probe scalar field. We conjecture that a similar threshold exists also in dynamical gravity in AdS. 
Since it is only in the limit of small\footnote{This $\epsilon$ is the (square-root) of the ratio of the two ``times" in the TTF formalism, see \cite{Buchel}.} $\epsilon$ that the TTF analysis of \cite{Buchel} is expected to hold, and since this corresponds precisely to the case where the amplitudes are forced to be small \cite{Buchel}, and since the instability identified by Bizon and Rostworoswki is expcted to be generic, we believe that an SST-like mechanism analogous to the one we found for the probe scalar should operate also for gravity. In a thought experiment  we may consider turning on gravity at a late time: if enough energy is transferred to localized modes in the position space, even if solely by scalar self-interaction, then eventually a black hole would form (see e.g. \cite{Zayas:2014ena} for connection of classical chaos and black hole formation). 

Our discussion opens up various intriguing questions, and we conclude this paper by mentioning two of them, which are currently under investigation.

\begin{itemize}
\item Is the total energy in the modes that is the relevant quntity for deciding the onset of thermalization or is it the maximum energy in any one mode? More pertinently, does the number of modes that one turns on affect the threshold at which thermalization sets in? 
How do things change when we slowly start increasing the number of modes. The results of \cite{Bizon} suggest that for smooth profiles, the instability is fairly generic. It will be interesting to understand the connection between the existance of this energy threshold and the threshold of black hole formation, a la Choptuik. Naively one might have thought that Coptuik's result is the holographic analog of the SST, but in AdS for {\em coherent} initial data, the observation of \cite{Bizon} is that black holes form for arbitrarily low amplitudes.




\item Is there something that can be said about the system analytically? Work along this lines is currently under progress. See \cite{Ponno}. 





\end{itemize}



\vspace{0.2in}
\acknowledgments
PB thanks Abhishek Dhar, Sumit Das and Oscar Dias for discussions; and YITP, Kyoto for hospitality during part of this work. CK thanks Ben Craps, Alex Buchel, Umut Gursoy, Sachin Vaidya and Leopoldo Pando Zayas for discussions/correspondence; and the Spinoza Institute (Utrecht) and ETH (Zurich) for hospitality during part of this work.

\bibliographystyle{JHEP}
\bibliography{sstPallab1-Aug1}

\end{document}